\documentclass[
preprint,
superscriptaddress,
showpacs,preprintnumbers,
 amsmath,amssymb,
 aps,
 prl
]{revtex4-1}

\usepackage[utf8]{inputenc} 
\usepackage{siunitx} 

\usepackage{graphicx}
\usepackage{dcolumn}
\usepackage{bm}

\begin{document}


\title{Evidence of large polarons in photoemission band mapping of the perovskite semiconductor CsPbBr$_3$}


\author{M. Puppin}%
\affiliation{Laboratoire de Spectroscopie Ultrarapide and Lausanne Centre for Ultrafast Science (LACUS), École Polytechnique Fédérale de Lausanne, ISIC, Station 6, CH-1015 Lausanne, Switzerland}%
\email{michele.puppin@epfl.ch}
\email{majed.chergui@epfl.ch}

\author{S. Polishchuk}%
\affiliation{Laboratoire de Spectroscopie Ultrarapide and Lausanne Centre for Ultrafast Science (LACUS), École Polytechnique Fédérale de Lausanne, ISIC, Station 6, CH-1015 Lausanne, Switzerland}%

\author{N. Colonna}%
\affiliation{Theory and Simulations of Materials (THEOS), and National Centre for Computational Design and Discovery of Novel Materials (MARVEL), École Polytechnique Fédérale de Lausanne, CH-1015 Lausanne, Switzerland}%

\author{A. Crepaldi}%
\affiliation{Institute of Physics and Lausanne Centre for Ultrafast Science (LACUS), École Polytechnique Fédérale de Lausanne, CH-1015 Lausanne, Switzerland}%

\author{D. N. Dirin}%
\author{O. Nazarenko}%
\affiliation{Laboratory of Inorganic Chemistry, Department of Chemistry and Applied Biosciences, ETH Z{\"u}rich, CH-8093 Z{\"u}rich, Switzerland}%
\affiliation{Laboratory for Thin Films and Photovoltaics, EMPA \\ Swiss Federal Laboratories for Materials Science and Technology, Überlandstrasse 129, CH-8600 D{\"u}bendorf, Switzerland}%

\author{R. De Gennaro}
\affiliation{Theory and Simulations of Materials (THEOS), and National Centre for Computational Design and Discovery of Novel Materials (MARVEL), École Polytechnique Fédérale de Lausanne, CH-1015 Lausanne, Switzerland}%

\author{G. Gatti}%
\author{S. Roth}%
\affiliation{Institute of Physics and Lausanne Centre for Ultrafast Science (LACUS), École Polytechnique Fédérale de Lausanne, CH-1015 Lausanne, Switzerland}%

\author{T. Barillot}%
\affiliation{Laboratoire de Spectroscopie Ultrarapide and Lausanne Centre for Ultrafast Science (LACUS), École Polytechnique Fédérale de Lausanne, ISIC, Station 6, CH-1015 Lausanne, Switzerland}%

\author{L. Poletto}%
\affiliation{National Research Council of Italy - Institute of Photonics and Nanotechnologies
(CNR-IFN), via Trasea 7, 35131 Padova, Italy}%

\author{R. P. Xian}%
\affiliation{Fritz-Haber-Institut der Max-Planck-Gesellschaft, Faradayweg 4-6, 14195 Berlin, Germany}%
\author{L. Rettig}%
\author{M. Wolf}%
\author{R. Ernstorfer}%
\affiliation{Fritz-Haber-Institut der Max-Planck-Gesellschaft, Faradayweg 4-6, 14195 Berlin, Germany}%

\author{M. V. Kovalenko}%
\affiliation{Laboratory of Inorganic Chemistry, Department of Chemistry and Applied Biosciences, ETH Z{\"u}rich, CH-8093 Z{\"u}rich, Switzerland}%
\affiliation{Laboratory for Thin Films and Photovoltaics, Empa \\ Swiss Federal Laboratories for Materials Science and Technology, Überlandstrasse 129, CH-8600 D{\"u}bendorf, Switzerland}%

\author{N. Marzari}%
\affiliation{Theory and Simulations of Materials (THEOS), and National Centre for Computational Design and Discovery of Novel Materials (MARVEL), École Polytechnique Fédérale de Lausanne, CH-1015 Lausanne, Switzerland}%

\author{M. Grioni}%
\affiliation{Institute of Physics and Lausanne Centre for Ultrafast Science (LACUS), École Polytechnique Fédérale de Lausanne, CH-1015 Lausanne, Switzerland}%

\author{M. Chergui}%
\affiliation{Laboratoire de Spectroscopie Ultrarapide and Lausanne Centre for Ultrafast Science (LACUS), École Polytechnique Fédérale de Lausanne, ISIC, Station 6, CH-1015 Lausanne, Switzerland}

\date{\today}

\begin{abstract}
Lead-halide perovskite (LHP) semiconductors are emergent optoelectronic materials with outstanding transport properties which are not yet fully understood. We find signatures of large polaron formation in the electronic structure of the inorganic LHP CsPbBr$_3$ by means of angle-resolved photoelectron spectroscopy. The experimental valence band dispersion shows a hole effective mass $0.26\pm0.02\,\,m_e$, 50$\,$\% heavier than the bare mass m$_0$=0.17 m$_e$ predicted by density functional theory. Calculations of electron-phonon coupling indicate that phonon dressing of the carriers mainly occurs via distortions of the Pb-Br bond with a Fr\"ohlich coupling parameter $\alpha=1.82$. A good agreement with our experimental data is obtained within the Feynmann polaron model, validating a viable theorical method to predict the carrier effective mass of LHPs ab-initio. \end{abstract}


\pacs{Valid PACS appear here}
\maketitle


\hyphenpenalty=10000
\exhyphenpenalty=2500
\tolerance=1500

\raggedbottom
Hybrid organic-inorganic and inorganic lead-halide perovskites (LHP) rival conventional semiconductors in multiple optoelectronic applications.  LHP-based solar cells have established energy conversion efficiencies approaching 25\% \cite{greenefficiency2019}; light-emitting devices \cite{quan_perovskites_2018} and lasers \cite{yakunin_low-threshold_2015} are gaining considerable interest thanks to high luminescence quantum efficiency \cite{johnston_hybrid_2016}. The carrier diffusion length is exceptionally long in LHPs, reaching up to several micrometers \cite{herz_charge-carrier_2017,manser2016intriguing}. This property results from long carrier lifetimes, rather than from the carrier mobility \cite{egger2018}. While theory predicts small effective masses \cite{becker_bright_2018,kang_intrinsic_2018,chen_structural_2018} ($\approx\,$0.1$\,$-$\,$0.2 $m_e$, where $m_e$ is the free electron mass), the reported mobilities are orders of magnitude lower than in conventional inorganic semiconductors \cite{egger2018,He2018}. 
The microscopic mechanism underlying this unusual combination of transport properties is possibly the interplay between carriers and the ionic perovskite lattice \cite{egger2018,miyata_large_2017}. In a polar crystal, longitudinal-optical (LO) phonon modes have a sizable long-range interaction with charge carriers, resulting in the formation of so-called Fr{\"o}hlich polarons \cite{fro1950}. The polaron, heavier than a bare carrier, has a reduced mobility, compatible with the observed transport properties \cite{zhu_charge_2015,miyata_large_2017}. In particular, the screening of the Coulomb potential is modified in the case of polarons, purportedly explaining the observed carrier lifetimes \cite{zhu_charge_2015,zhu2016screening}. 

The optical properties of different LHPs are known to critically depend on the details of the lead-halide bond angles \cite{stoumpos2015renaissance}, highlighting the importance of carrier-lattice coupling also in the photophysics of LHPs. The presence of polaron quasi-particles was indeed already proposed to model the results of several optical studies \cite{juarez-perez_photoinduced_2014,zhu_charge_2015,zhu2016screening}.

In this letter we report on experimental evidence of polaron formation by measuring its fingerprint in the electronic structure.
We concentrate on the prototypical inorganic LHP CsPbBr$_{3}$ which has lately attracted interest for applications, due to better thermal and radiation stability compared to hybrid organic-inorganic LHPs \cite{turren-cruz_methylammonium-free_2018, song_probing_2018,kulbak_cesium_2016,liang_all-inorganic_2017,calisi_composition-dependent_2018}. The momentum-resolved electronic structure of CsPbBr$_{3}$ is determined by angle-resolved photoelectron spectroscopy (ARPES) and compared with ab-initio density functional theory (DFT). Our ARPES data provide a direct measurement of the hole effective mass ($m_{exp}$) in CsPbBr$_{3}$. The experiment reveals a mass enhancement of 50\% compared to theory, which we attribute to electron-phonon coupling. Ab-initio simulations of electron-phonon interaction show that Pb-Br stretching modes dominate the interaction. Furthermore, our calculations provide a Fr\"ohlich coupling parameter $\alpha=1.82$, which indicate that carriers form large polarons and predict a mass renormalization in good agreement with experimental data.

\begin{figure*}[htp]
\includegraphics[width=1.0\columnwidth]{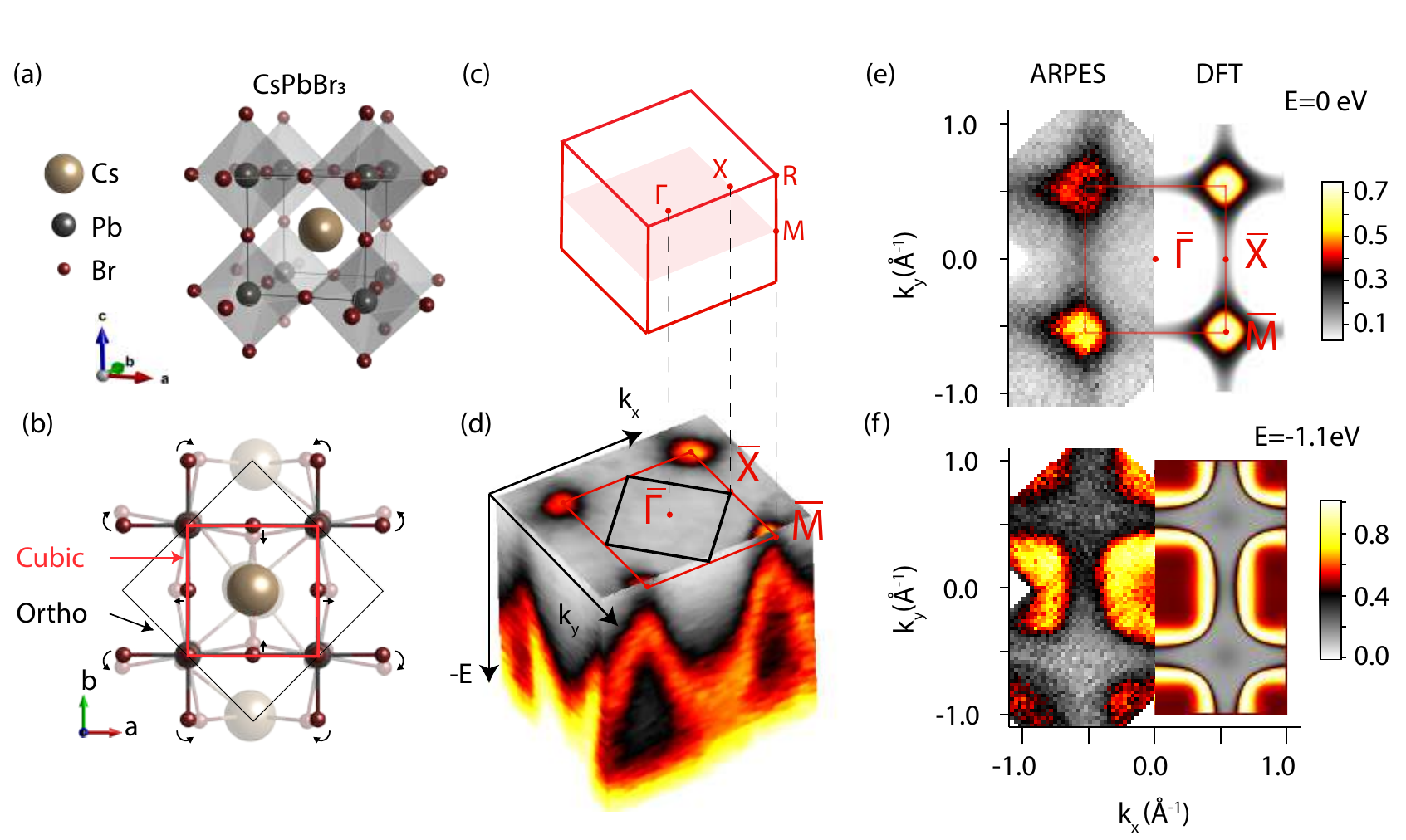}
\caption{\label{intro} Schematic structure of CsPbBr$_3$: (a) crystal cell; [PbBr$_{6}$]$^{4-}$ octahedra are indicated as shaded-gray surfaces, Pb$^{2+}$ ions are indicated in black, Br$^{-}$ ions in red and the Cs$^{+}$ cation in gold; (b) the orthorhombic lattice distortion (semitransparent lines) is compared to the parent cubic lattice (full lines); (c) Three-dimensional Brillouin zone of the cubic crystal lattice; (d) VB ARPES intensity as a function of energy, E and in-plane momentum wavevectors, $k_{x}$ and $k_{y}$. The cubic and orthorhombic unit cells are indicated in red and black, respectively. (e-f) Constant energy cuts of the ARPES intensity compared with DFT calculations at the VBM (E=0 eV, (e)) and  below the VBM (E=-1.1 eV. (f)).}.  
\end{figure*}
The high temperature (T $>$ 130 $^{\circ}$C) lattice structure of CsPbBr$_3$ [Fig. \ref{intro} (a)] consists of a cubic arrangement of corner-sharing [PbBr$_{6}$]$^{4-}$ octahedra, where a Pb$^{2+}$ ion is surrounded by four Br$^{-}$ ions. This backbone cages the Cs$^{+}$ cation. 
The corresponding reciprocal space Brillouin zone is depicted in Fig. \ref{intro} (c).
Upon cooling below 130$^{\circ}$C, the system first undergoes a structural phase transition to a tetragonal phase, finally followed at 88 $^{\circ}$C by a transition to an orthorhombic phase, which is the stable room-temperature lattice structure. The structural phase transitions cause the PbBr$_6$ octahedra to reorient, reducing the crystal symmetry \cite{stoumpos_crystal_2013}. The orthorhombic phase is compared to the undistorted cubic phase in Fig. \ref{intro} (b), showing its larger real-space primitive cell and the octahedra's canting angle of approximately 10$^{\circ}$ \cite{He2018}.

High-quality single crystals of CsPbBr$_3$ were grown from liquid solution using an inverse temperature crystallization method \cite{dirin_solution-grown_2016}. 
The CsPbBr$_3$ crystals were cleaved in-situ under ultra-high vacuum conditions. 
ARPES experiments were performed using extreme ultraviolet radiation from a high-harmonic laser source with a tunable photon energy between 20 and 40~eV \cite{ojeda_harmonium:_2015,crepaldi_time-resolved_2017}. All data were collected at room temperature, in the orthorhombic phase of CsPbBr$_3$, as confirmed by X-ray diffraction \cite{note:SI}. 
To rationalize the experimental results, we performed ab-initio calculation using the \textsc{Quantum ESPRESSO}
distribution~\cite{giannozzi_quantum_2009, giannozzi_advanced_2017}.
The electronic structure was obtained at the generalized Kohn-Sham level
using the hybrid functional scheme proposed by Heyd, Scuseria and 
Ernzerhof~\cite{heyd_hybrid_2003,heyd_erratum_2006} (HSE) for the 
exchange and correlation energy functional. The electron-phonon interaction 
was accounted for within the Fr\"ohlich model~\cite{frohlich_electrons_1954}
with parameters obtained averaging the ab-initio Fr\"ohlich 
vertex~\cite{vogl_microscopic_1976, verdi_frohlich_2015}.
Further details concerning the experimental methods and the DFT calculations are given as supplemental informations \cite{note:SI}. 

The valence band (VB) photoemission intensity distribution is plotted as a function of energy and in-plane momentum wavevectors in Fig.~\ref{intro}~(d), for a photon energy of 37~eV. Four valence band maxima (VBM) are clearly resolved, following the periodicity of the surface-projected Brillouin zone (SBZ) of the cubic phase. Figure \ref{intro} (e) and (f) shows two cuts at constant energy of the three-dimensional ARPES intensity distribution, at the VBM and 1.1 eV below the VBM. The energy zero was set at the VBM, determined from the energy of the peak maximum. The experimental data follows the square symmetry throughout the measured energy range, and the VBM are located at the four corners ($\overline{M}$ point) of the SBZ. 

\begin{figure}[ht]
\includegraphics[width=0.5\columnwidth]{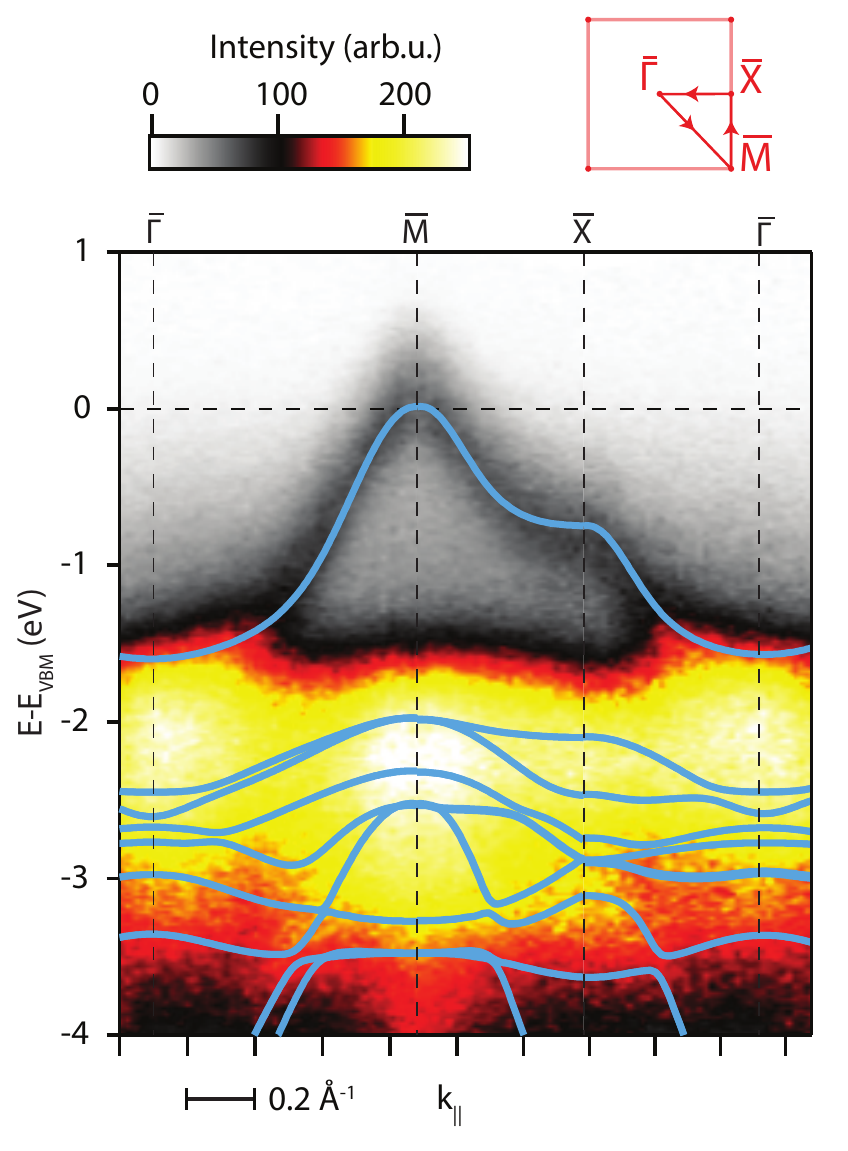}
\caption{\label{bands} Photoemission intensity as a function of energy and parallel momentum, along the path $\overline{\Gamma}-\overline{M}-\overline{X}-\overline{\Gamma}$, indicated in the top right panel. The computed DFT bands (cyan) for the cubic phase are overlaid on the data.}
\end{figure}

This is at odd with DFT calculations for the orthorhombic phase, which predicts an additional (back-folded) VBM at the $\overline{\Gamma}$ point \cite{note:SI}. To exclude matrix element effects and dispersion in the direction orthogonal to the sample surface ($k_{\perp}$), we performed energy- and polarization-dependent ARPES measurements \cite{note:SI}, which reveal no signature of an additional VBM at the $\overline{\Gamma}$ point. The observation of a larger k-space periodicity is not compatible with the scenario of a surface reconstruction. The additional potential associated with a periodic lattice distortion, such as that occurring in the orthorhombic phase, generally manifests itself with the appearance of back-folded bands and gaps opening at the novel Bragg planes. However, the spectral weight transfer to the novel bands is proportional to the strength of the perturbing potential and often hardly observable~\cite{Voit2000ARPESperiodic}, e.g. for the methylammonium lead triiodide perovskite (MAPbI$_{3}$)~\cite{lee_first_2017,yang_band_2018}, where no signatures of back-folded orthorhombic bands were observed by ARPES, despite a clear orthorhombic diffraction pattern.

The absence of a significant spectral weight transfer to the orthorhombic periodicity implies that the bands calculated for the cubic phase overlap well with the ARPES spectra. The data are compared to theoretical results for the cubic phase on the right half of each panel of Fig. \ref{intro} (e) and (f). The finite experimental momentum resolution in $k_{\perp}$, due to the short photoelectron mean free path, is accounted for by integrating the DFT bands over a range of 0.1~\AA$^{-1}$ along the $k_{\perp}$ direction, corresponding an estimated escape depth of 5~\AA\ \cite{komesu_surface_2016}.
The material's band structure has been investigated as a function of the photon energy, and Fig. \ref{bands} shows the result for 33.5 eV, which is found to be close to the bulk R point \cite{note:SI}. The data correspond to the band dispersion along three high-symmetry directions ($\overline{\Gamma}-\overline{M}$, $\overline{X}-\overline{M}$ and $\overline{\Gamma}-\overline{X}$) and are compared with the calculated bands in the bulk X-M-R plane. The upper valence band disperses for approximately 1.5~eV below the VBM, before reaching a deeper valence manifold, where bands are not individually resolved. 
Simulated element-projected partial density of states reveals that the highest-energy VB is mainly composed of Pb~6s and Br~4p orbitals derived from the PbBr$_6$ octahedra, in accord with previous calculations \cite{yettapu2016}.

\begin{figure}[ht]
\includegraphics[width=0.5\columnwidth]{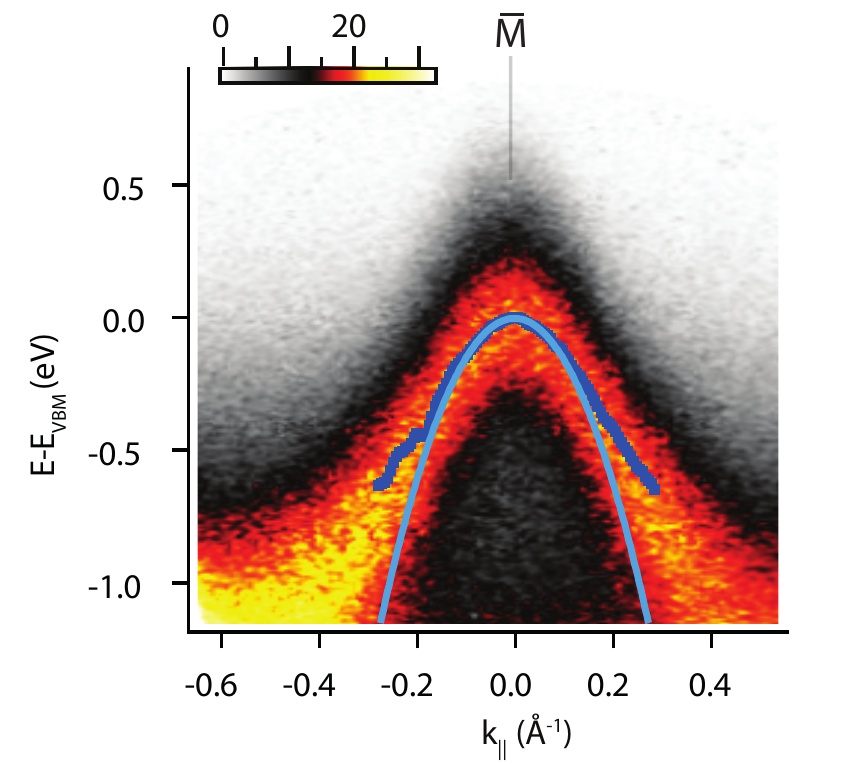}
\caption{\label{fitdescription} ARPES intensity as a function of energy and parallel momentum showing the VBM along the $\overline{\Gamma}-\overline{M}-\overline{\Gamma}$ direction . The fitted band maxima are indicated as blue points. Cyan curve: parabolic band fitted around the band maximum.}
\end{figure}

Although in the room-temperature orthorhombic phase the ARPES spectral weight follows qualitatively the DFT bands for the cubic phase, the band dispersion is modified by the structural distortion. In fact, a comparison between DFT calculations of the two phases reveals that the effective mass computed for the orthorhombic phase is 0.17 m$_e$, higher than the cubic phase mass of 0.15 m$_e$ \cite{note:SI}. To determine the experimental hole effective mass, we turn to a quantitative analysis of the upper valence band dispersion which we compare with ab-initio calculations for the orthorhombic structure. ARPES data along the $\overline{\Gamma}-\overline{M}$ direction are shown in Fig. \ref{fitdescription}. The VB energy distribution curves are well fitted by a Gaussian line shape whose width (which is not resolution-limited) is likely determined by thermal broadening with possible contributions from disorder and orthogonal momentum dispersion. To determine $m_{exp}$, the valence band was fitted with a parabolic dispersion around the band maximum, until convergence was observed \cite{note:SI}, the corresponding best fit is shown in Fig. \ref{fitdescription}. The obtained value $m_{h}=0.26\pm0.02\,m_e$ is in good agreement with optical measurements on CsPbBr$_{3}$ \cite{yang_impact_2017}, where a reduced exciton mass of $m_{exc}=0.126\,m_e$ was deduced, if one assumes balanced electron and hole effective masses, which appears justified by our DFT calculations.

\begin{figure*}[ht]
\includegraphics[width=0.75\columnwidth]{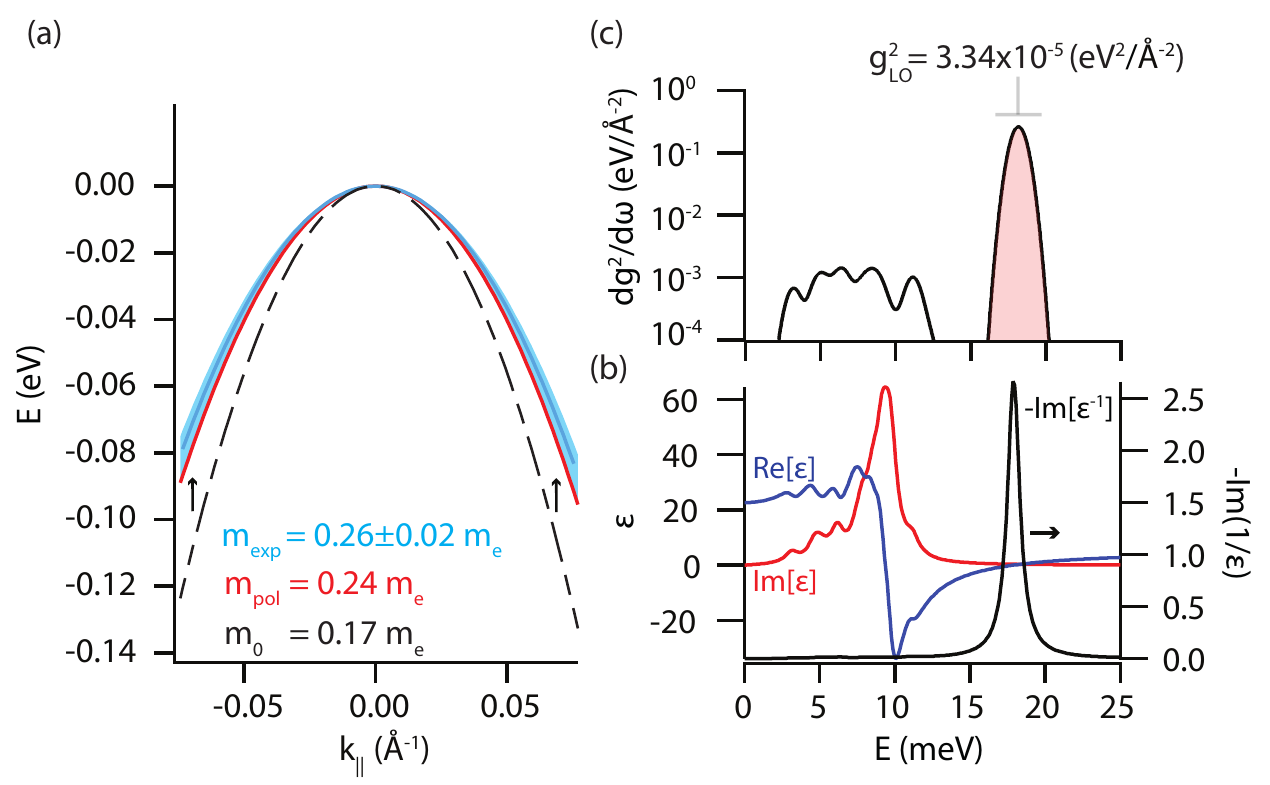}
\caption{\label{fig:theory} (a) Comparison between the experimental dispersion (m$_{exp}$, cyan line, the shaded area indicates the experimental uncertainty) and the theoretical effective mass m$_{0}$ computed from theory (black dot-dashed line). The renormalized mass including electron-phonon interaction m$_{pol}$ is plotted in red. (b) Computed dieletric function, real (blue line) and imaginary (red line) part are shown on the left-hand axis; the loss function $-Im(1/\epsilon)$ is plotted on the right-hand axis. (c) Logarithmic plot of the density of coupling $d(g^2)/d\omega$ to optical phonons \cite{note:SI}, the shaded area indicates the integration region for determining the coupling constant $g^2_{LO}$. $d(g^2)/d\omega$ was broadened by convolution with a Gaussian function (1.2 meV FWHM) for clarity.}
\end{figure*}

The effective mass calculated at the HSE level of theory for the orthorhombic phase ($m_{0}$) is compared to $m_{exp}$ in Fig. \ref{fig:theory} (a). Theory substantially underestimates $m_{exp}$, with an experimental mass enhancement of $\approx 50\%$, implying the presence of a mass renormalization mechanism. 
Comparison between HSE and G$_0$W$_0$ effective masses shows minor changes ($\approx \,$8\%), indicating that the hybrid HSE functional gives a reasonable description of the band structure \cite{note:SI}. These findings seems to rule out electronic correlation effects as the main reason for the mass enhancement observed.

An important mechanism, not accounted for by the DFT calculations and relevant for polar materials, is the interaction between electrons and longitudinal optical phonons. ARPES is sensitive to such many-body interactions, encoded in the single particle spectral function \cite{damascelli_probing_2004}. In particular, for polaronic systems, such interactions manifest themselves as a renormalization of the bare band dispersion and with the appearence of satellite peaks in the photoemission spectrum \cite{moser2013,wang_tailoring_2016}. The satellites appear on the low-energy side of the main quasi-particle peak, at an energy separation corresponding to the relevant longitudinal optical (LO) phonon mode. 
In CsPbBr$_{3}$ optical phonons have a energies $\leq\,$25 meV \cite{calistru1997identification,guo2017polar}, and replicas cannot be resolved within the experimental linewidth. In contrast, our analysis of the quasi-particle dispersion captures the effective mass renormalization, which we attribute to electron-phonon interaction. 

This interpretation is supported by recent theoretical predictions for CsPbBr$_{3}$ and related compounds, e.g. MAPbI$_{3}$, which exhibits the same lattice structure and similar phase diagram. Simulations of the electron-phonon interaction in MAPbI$_{3}$ predict a mass enhancement of $\approx\,$30$\,$\%, where the interaction is dominated by coupling with longitudinal optical phonon modes, the most important being the Pb-I stretching and bending modes, and the librational-translational modes of the methylammonium cation \cite{PhysRevLett.121.086402}. Since the latter modes are absent in the fully inorganic compound, we expect the largest contribution to arise from the Pb-Br bond. Simulations of hole addition into the CsPbBr$_{3}$ lattice were performed by Miyata et al. \cite{miyata_large_2017}, showing that the largest structural relaxation occurs on the Pb-Br bond and on the Pb-Br-Pb bond angle, resulting in a reduction of the canting angle of the PbBr$_6$ octahedra towards the undistorded cubic lattice. 

To validate this picture, we performed ab-initio calculations of the phonon bandstructure of orthorhombic CsPbBr$_{3}$ and of its dielectric function, reported in Fig. \ref{fig:theory} (b). To estimate the Fr\"ohlich interaction, we follow a method recently developed for polar semiconductors \cite{verdi_frohlich_2015,PhysRevLett.121.086402}. The Fr\"ohlich vertex, which represents the matrix element for electron scattering by long-wavelength longitudinal optical phonons, can be written~\cite{vogl_microscopic_1976, verdi_frohlich_2015} as: 
\begin{equation}\label{eq:g}
g_{\nu}(\mathbf{q}) = -i\frac{4\pi e^2}{\Omega} \sum_k \sqrt{\frac{\hbar}{2M_k\omega_{\mathbf{q}\nu}}} 
                \frac{ \hat{\mathbf{q}} \cdot Z^*_k \cdot \mathbf{e}_{k{\nu}}(\mathbf{q}) }{ \hat{\mathbf{q}} \cdot \varepsilon_{\infty} \cdot \hat{\mathbf{q}} }
\end{equation}
where e is the electron charge, $\Omega$ is the volume of the unit cell, $M_k$ the mass of the atom $k$, $Z^*_k$ the Born effective
charge tensor, $\varepsilon_{\infty}$ the high-frequency dielectric tensor, and $\omega_{\mathbf{q}\nu}$ and 
$\mathbf{e}_{k\nu}(\mathbf{q})$ the eigenvalue and eigenvector associated with the mode $\nu$ of momentum $\mathbf{q}$. To assess the relative importance of different phononic contributions in our calculations, the energy density of coupling $d(g^2)/d\omega$ \cite{note:SI} is plotted as a function of phonon energy in Fig. \ref{fig:theory} (c). The coupling is dominated by a maximum at an effective energy of $ \hbar\tilde{\omega}_{LO} =18.2$ meV, in the energy region of Pb-Br stretching modes \cite{miyata_large_2017}. The effective electron-phonon coupling to such modes is obtained integrating $d(g^2)/d\omega$ from 12 to 25~meV [see Figure \ref{fig:theory} (b)], resulting into $\tilde{g}^2_{LO}=3.34\times10^{-5}$ eV$^2$/\AA$^{-2}$. Our calculation reveals that the coupling to the Pb-Br stretching modes is two orders of magnitude stronger compared to modes appearing in the energy range between 2 and 13 meV in Figure \ref{fig:theory} (c), which can be associated with coupled stretching-bending modes of Pb-Br \cite{miyata_large_2017}. 

Following these calculations, we proceed to estimate the mass renormalization from the Fr{\"o}hlich model \cite{frohlich_electrons_1954}, valid for a parabolic band dispersion and coupling to a single dispersionless LO phonon mode. In this limit, it can be shown that the coupling matrix elements $g_{\nu}(\mathbf{q})$ reduces to the well-known Fr{\"o}hlich coupling matrix elements \cite{verdi_frohlich_2015}.
The dimensionless Fr{\"o}hlich coupling parameter, $\alpha$, can be expressed in term of the ab-initio effective coupling strength $\tilde{g}^2_{LO}$ as:
\begin{equation}\label{eq:alpha}
 \alpha = \frac{\Omega}{4\pi e^2} \frac{ \tilde{g}^2_{LO} }{(\hbar\tilde{\omega}_{LO})^2} 
                \left( \frac{2m_0\tilde{\omega}_{LO}}{\hbar} \right)^{1/2}.
\end{equation}
with $m_0$ the bare effective mass. We obtain $\alpha=1.82$, which fall into the small to intermediate coupling regime. In this regime, the Feynman polaron model provides a good approximation for the quasi-particle mass \cite{Feynman1962,DevreeseREVPOL,PhysRevLett.121.086402}:
\begin{equation}\label{eq:meff}
m_{pol} = m_{0}(1+\frac{\alpha}{6} + 0.025\alpha^2 +.. ),
\end{equation}
Here $m_{pol}$ is the renormalized polaron mass, and $m_{0}$ is the bare quasi-particle mass extracted from our DFT calculations. The resulting $m_{pol}=0.24\,m_e$ is compared to the experimental result in Fig. \ref{fig:theory}. 
The result, in agreement with experiment within the experimental uncertainty, indicates that our model captures the main physics behind the hole quasi-particle dressing. The coupling of carriers to the Pb-Br bond modes, might play an important role also on the optical properties of LHPs, which critically depend on the Pb-Br-Pb bond angle \cite{stoumpos2015renaissance}.
Within the Feynman model, it is also possible to estimate the polaron binding energy and radius to be 33.5~meV and 58~\AA, respectively. Thus, the polaron resulting from an excess hole in CsPbBr$_{3}$ single-crystals is large, extending over several lattice unit cells. Interestingly, in the case of CsPbBr$_{3}$ nanocrystals, signatures of hole self-trapping were reported \cite{santomauro2017localized}, suggesting that the electron-phonon interaction in LHPs nanostructures may be altered \cite{neukirch2016polaron,iaru2017strong}. 
The adopted theoretical method can be readily generalized to multiple coupled LO phonon modes \cite{PhysRevLett.121.086402}, as in the case of hybrid organic-inorganic LHPs. Therefore, we expect it to be capable of predicting the carrier effective masses in the whole family of LHPs.

In conclusion, our work provides the first experimental reference for the momentum-resolved electronic structure of CsPbBr$_{3}$ in the orthorhombic phase. Fits of the electronic dispersion provide an experimental value for the effective mass $m_{exp}=0.26\pm0.02\,m_e$, which we found to exceed the theoretical result of $m_{0}=0.17\,m_e$. The observed mass renormalization is ascribed to electron-phonon interaction dominated by Pb-Br stretching modes, responsible for the formation of large Fr{\"o}hlich polarons. Ab-initio calculations are in quantitative agreement with the experiment demonstrating that the employed theoretical method can correctly predict the carrier effective mass of LHPs from first principles. Our findings provide direct experimental evidence in the electronic structure that charge carriers in single-crystalline LHPs form large polarons and that the corresponding modification to the microscopic scattering rates must be taken into account to explain the exceptional transport properties of LHPs.  

This work was supported by the ERC grant DYNAMOX, the Max-Planck-EPFL Center of Molecular Nanoscience and Technology and the Swiss National Science Foundation via the NCCR’s MUST and MARVEL and Grant No. 200021-179139.

M.P., S.P, N.C and A.C. wrote the manuscript; S.P. and M.P. analyzed the experimental data; S.P., M.P, R.P.X. and L.R. acquired the ARPES data; D.N.D., O.N. and M.V.K. grew and characterized the samples; N.C., R.d.G. and N.M. performed the numerical calculations; M.P., S.P., A.C., G.G., S.R., T.B., L.P. provided technical support during the experiments; M.C. conceived and supervised the project, all authors contributed to the discussion and provided critical feedback to the manuscript.



\pagebreak

\onecolumngrid
\begin{center}
  \textbf{\large Evidence of large polarons in photoemission band mapping of the perovskite semiconductor CsPbBr$_3$\\Supplemental information}\\[.2cm]
(Dated: \today)\\[1cm]
\end{center}

\setcounter{equation}{0}
\setcounter{figure}{0}
\setcounter{table}{0}
\setcounter{page}{1}
\renewcommand{\theequation}{S\arabic{equation}}
\renewcommand{\thefigure}{S\arabic{figure}}
\section{Experimental methods}

\subsection{ARPES experimental details.}

\begin{figure*}[ht]
\includegraphics[width=0.75\columnwidth]{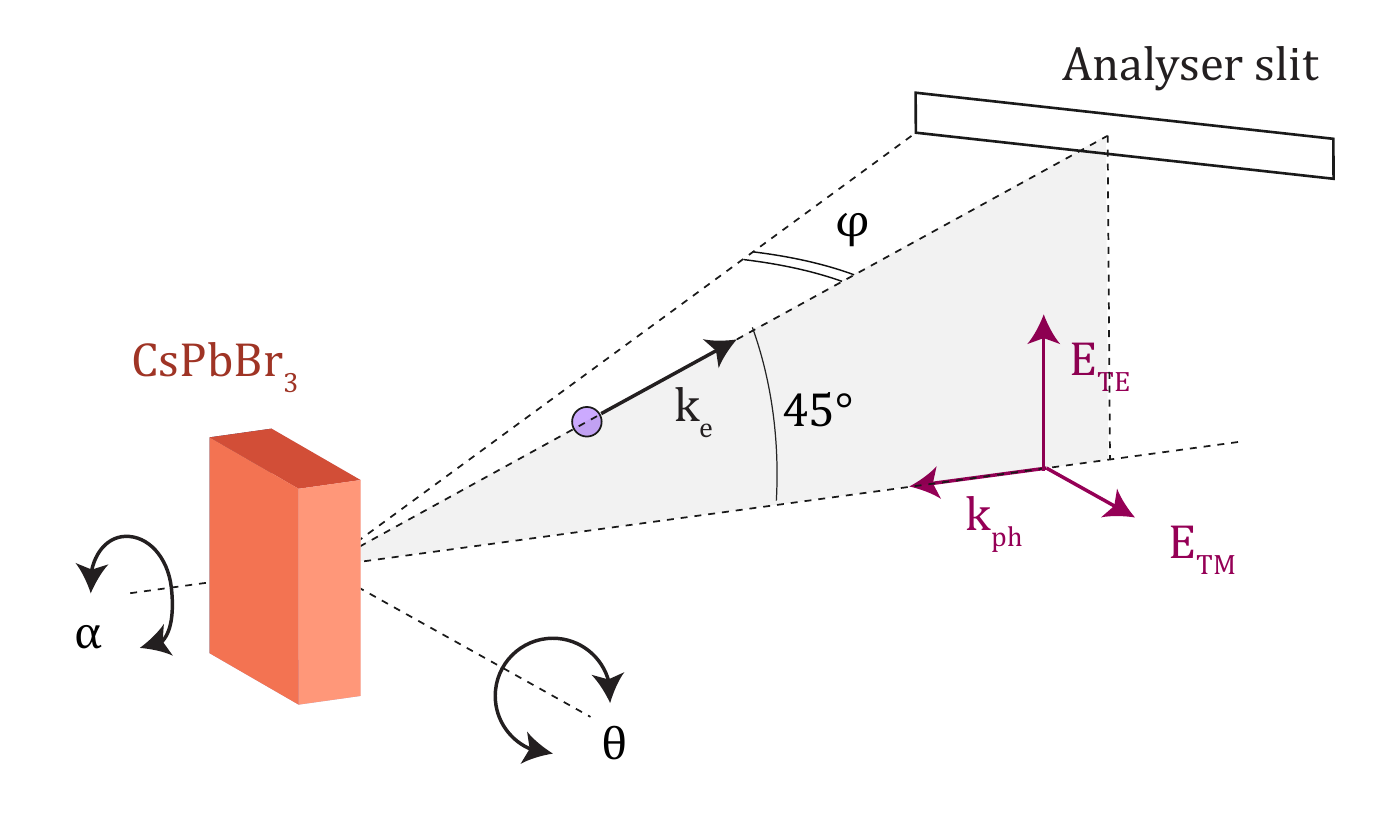}
\caption{\label{fig:ARPESgeometry} ARPES experimental geometry, the scattering plane, relative to the analyser slit is indicated in gray. The sample is installed on a manipulator where the azimutal angle $\alpha$ and polar angle $\theta$ can be adjusted.}  
\end{figure*}

The geometry of the ARPES experiment is sketched in Fig. \ref{fig:ARPESgeometry}. The XUV beam is linearly polarized and the polarization can be continuously rotated between transverse electric (TE) and transverse magnetic (TM). For all the data shown in the main text, the polarization was fixed in the TE configuration, with the electric field vector $E_{TE}$ in the scattering plane, which is defined by the wavevector of the incoming XUV photon and the wavevector of photoelectrons emitted toward the center of the analyser's slit (gray plane in Fig. \ref{fig:ARPESgeometry}). The analyser is fixed at an angle of 45$^{\circ}$ from the incoming photons. The analyzer slit is orthogonal to the scattering plane and the analyser has an angular acceptance of $\pm 15^{\circ}$. By changing the azimuthal angle $\alpha$ the sample was oriented along the ($\overline{\Gamma M}$) direction. The whole surface BZ was sampled by varying the polar angle $\Theta$ (between the surface normal and the analyzer) using a motorized manipulator.
The XUV radiation is produced by high-harmonic generation of a femtosecond laser in Argon, followed by a monochromator tunable between 20 and 40 eV. The spot size on the sample has a 100 $\mu$m diameter, with a flux on the order of 10$^{10}$.
For the scan displayed in Fig. 1 and 2 of the article, the estimated energy resolution originating from the finite bandwidth of the XUV source is $\Delta$E$_s\approx200$ meV at 37 eV, while the resolution of the photoelectron analyser is approximately $\Delta$E$_s\approx150$ meV. The experimental energy resolution can be estimated as $\Delta$E$_{exp}\approx250$ meV.
In the case of Fig. 3 of the main text the energy resolution of the photoelectron analyzer was set not worsen significantly the light source resolution, which was better than $\approx150$ meV at 33.5 eV.

\subsection{$\overline{\Gamma}$ point evolution with photon energy and light polarization.}
\begin{figure*}[ht]
\includegraphics[width=0.75\columnwidth]{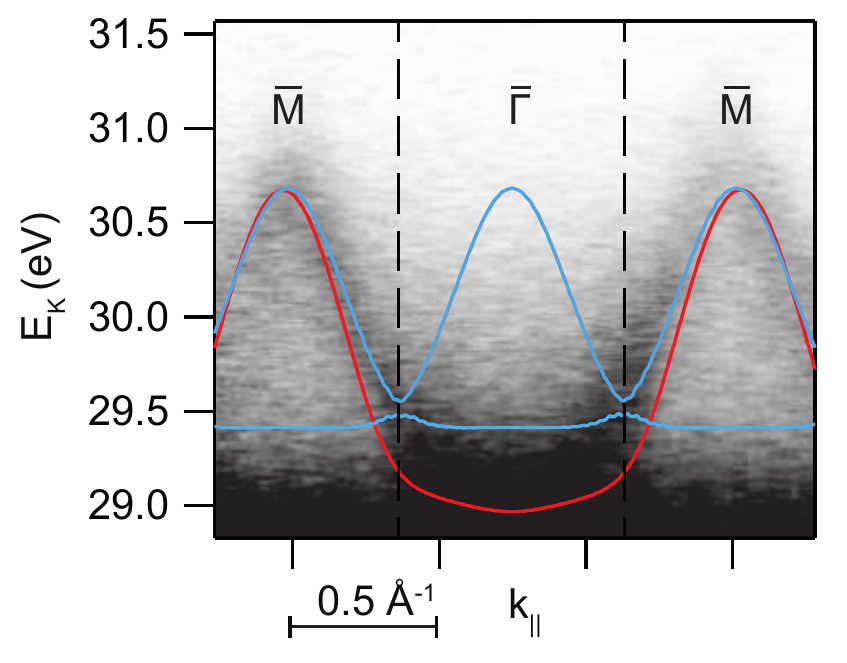}
\caption{\label{fig:CUBEORTHO} Experimental ARPES data compared along the $\overline{M}-\overline{\Gamma}-\overline{M}$ direction, HSE-DFT calculations are reported for the cubic phase (red line) and the orthorhombic phase (blue lines). The BZ border for the orthorhombic phase is marked by vertical dashed lines}  
\end{figure*}

\begin{figure*}[ht]
\includegraphics[width=1.0\columnwidth]{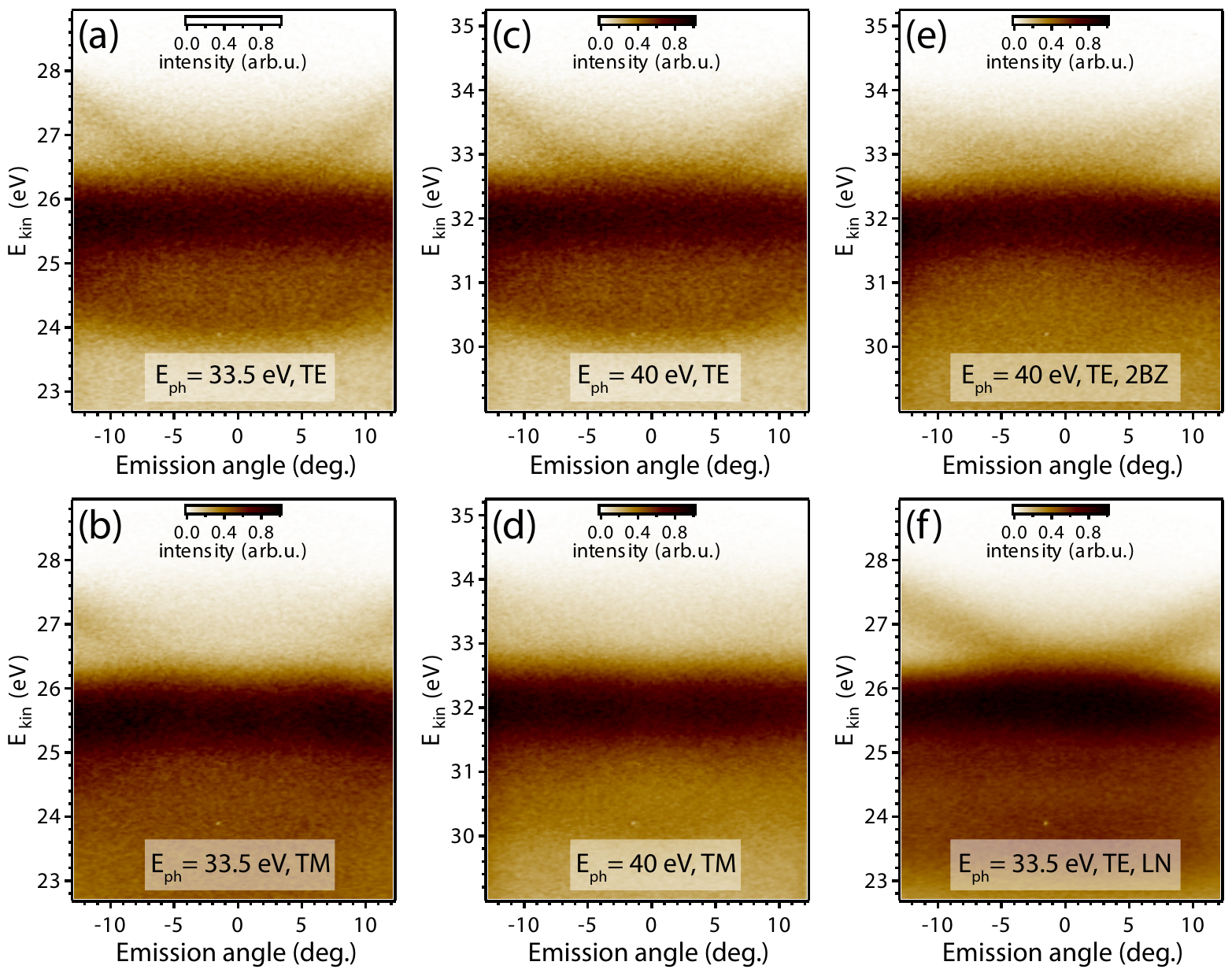}
\caption{\label{fig:GAMMA} ARPES images across the $\overline{M \Gamma M}$ direction showing the absence of photoemission intensity developing at the $\overline{\Gamma}$ point for different photon energy, probe polarization, temperature, and observation in the second Brillouin zone (2BZ): (a), (b) for the TE and TM polarized probe, respectively, at 33.5 eV photon energy; (c), (d) - same at 40 eV photon energy; (e) – at the $\overline{\Gamma}$ point in the 2BZ, with TE polarized probe; (d) at liquid nitrogen temperature (LN $\approx \,$77 K), with TE polarized probe. } 
\end{figure*}

The experimental data collected along the $\overline{M}-\overline{\Gamma}-\overline{M}$ direction is reported in Fig. \ref{fig:CUBEORTHO}, together with HSE-DFT calculations in both the cubic and orthorhombic phase. For the orthorhombic phase, theory predicts an additional valence band maximum (VBM) at the $\overline{\Gamma}$ point, which is not observed in our data.
To discard the possibility of a matrix element effect or an incorrect location in reciprocal space, we performed polarization- and energy-resolved measurements. Selected ARPES spectra are shown along the $\overline{M}-\overline{\Gamma}-\overline{M}$ direction in Fig. \ref{fig:GAMMA}. In both TE and TM polarizations and also in the second BZ, no band was resolved, which makes a strong-matrix element effect unlikely. Furthermore the energy was varied at normal emission condition between 20 and 40~eV, without the appearance of a VBM at the $\overline{\Gamma}$ point. These findings support the interpretation that the spectral intensity transferred to the backfolded VB at the $\overline{\Gamma}$ point is below our detection sensitivity. 

\subsection{Experimental determination of effective masses}

\begin{figure*}[ht]
\includegraphics[width=1.0\columnwidth]{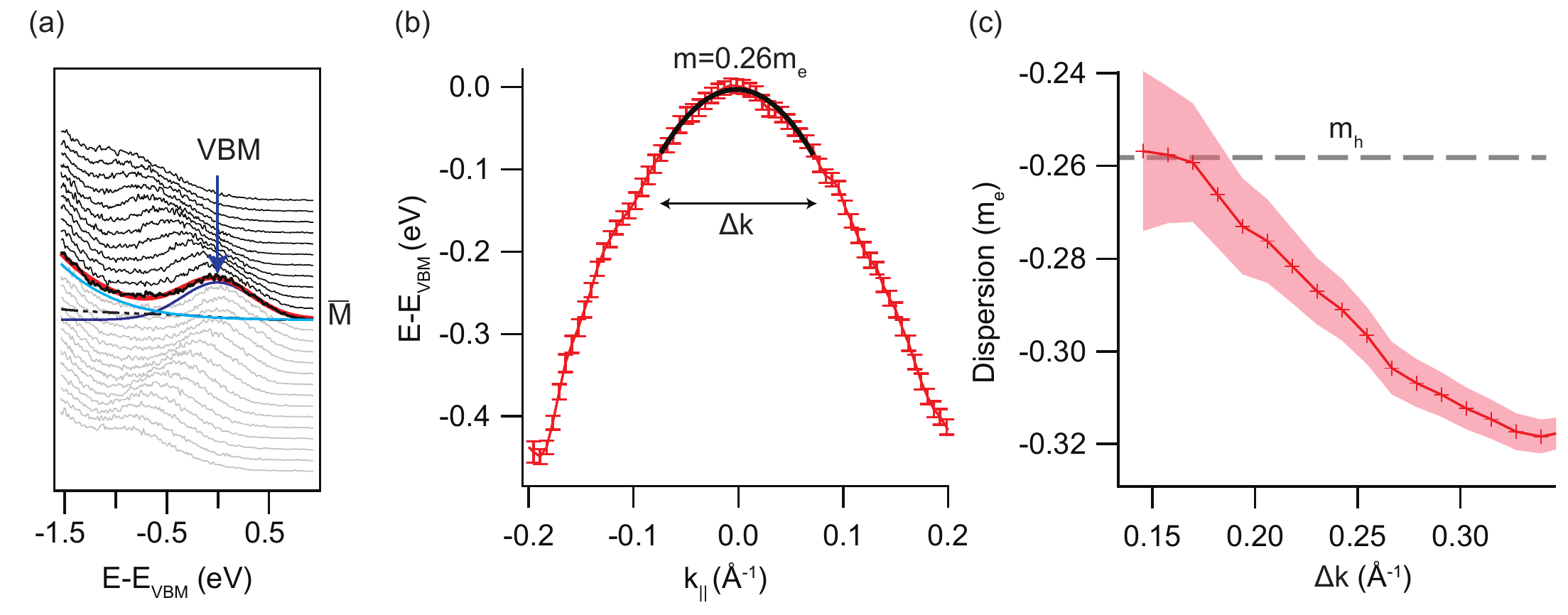}
\caption{\label{fig:SIMASS} (a) Energy distribution curves along the $\overline{M}-\overline{\Gamma}-\overline{M}$ direction, offset vertically for different values of $k_{\parallel}$. The fit (red line) is illustrated for the curve at the $\overline{M}$ point: in blue and cyan, Gaussian peaks, fitting the valence band maximum and a lower lying band, respectively. The dash-dotted line is an integral Shirley-type background. (b) .(c).}  
\end{figure*}

The VBM fit function is illustrated in Fig. \ref{fig:SIMASS} (a): the peak is well fitted by a Gaussian function (blue curve). A Shirley-type integral background was used (black dash-dot curve), with a coefficient taken to match the offset in a lower-lying portion of the spectrum, free of spectral features. The tail of the lower-lying valence band was approximated by a second Gaussian peak at lower binding energy (cyan curve). The resulting fit function (red curve) well approximate the data in a region of $\pm$0.2 \AA$^{-1}$ around the $\overline{M}$.
The position of the VBM peak is shown in in Fig. \ref{fig:SIMASS} (b). A parabolic curve was fitted for increasingly narrower regions $k_{max} \pm \Delta k/2$ around the maximum. The corresponding effective mass, together with the fit error (plus or minus one standard deviation), are indicated in Fig. \ref{fig:SIMASS} (c). The parabolic model fits well the data below $\Delta k=0.2$\AA$^{-1}$, the region was decreased symmetrically by one data point below and after the maximum, until the variation of the fit value was less than 1\% for two successive steps, the chosen condition for fit convergence. 

\subsection{Determination of k$_{\perp}$.}
The finite photoelectron escape depth determines an uncertainty in the value of the electron momentum in the direction orthogonal to the sample surface k$_{\perp}$. In the case of lead halide perovskites, the inelastic mean free path (IMFP) was estimated from the universal curve taking into account the presence of heavy Pb and Br atoms in reference \cite{komesu_surface_2016_SI}. For an IMFP of 5 \AA, the FWHM width the k$_{\perp}$ distribution is $\approx\,$0.1 \AA$^{-1}$, which corresponds to about $\pm$20\% of the M to R distance in reciprocal space. Under these conditions, ARPES still provides reasonable k$_{\perp}$ selectivity \cite{STROCOV200365_SI}.

\begin{figure*}[ht]
\includegraphics[width=1.0\columnwidth]{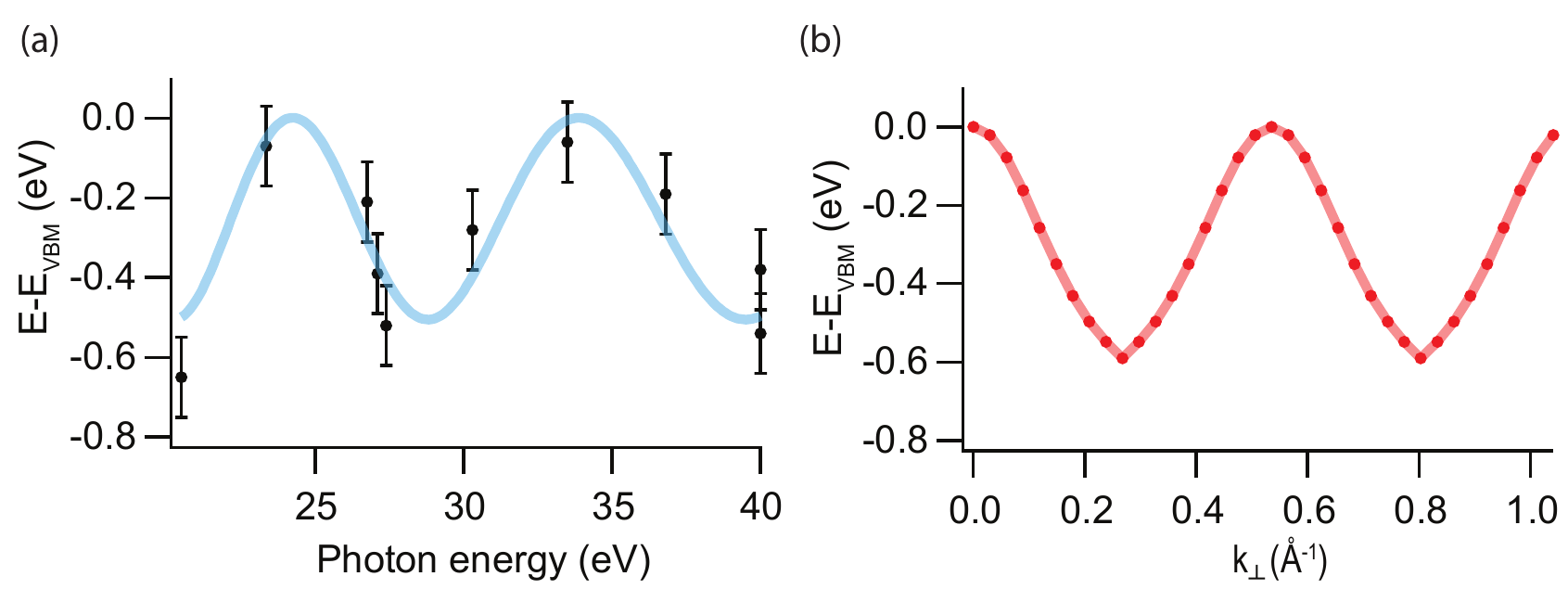}
\caption{\label{fig:kperpBE} (a) Measured dispersion of the VBM at the $\overline{M}$ point as a function of the photon energy, the cyan curve is a sinusoidal curve fitted to the data as a function of $k_{\perp}(\hbar \omega)$ (b) Theoretical energy dispersion extracted at the $\overline{M}$ point in the orthorhombic phase}  
\end{figure*}

To determine k$_{\perp}$ we follow the evolution of the $\overline{M}$ point as a function of energy. The fitted VBM is shown in Fig. \ref{fig:kperpBE} (a), the corresponding theoretical dispersion is shown in Fig. \ref{fig:kperpBE} (b). We used a free-electron final state model:
\begin{equation}
\label{eq:kperp}
 k_{\perp}= \frac{\sqrt{2m_e}}{\hbar} \sqrt{E_K \cos^2(\theta)  +V_0 }
\end{equation}
and fitted the band dispersion with a sinusoidal function $E_0 \sin(2 \pi k_{\perp}(\hbar \omega)/k_0 +\phi  )$, whose periodicity $k_0$ was fixed to match the known lattice parameter, and those phase $\phi$ was fixed to match the theoretical energy oscillation phase. We obtain a value of $V_0=0.7\pm0.7$ eV, from which we obtain a value $k_{\perp}=0.49 \pm 0.04$ \AA$^{-1}$ for the $\overline{M}$ point measured at a photon energy of 33.5 eV, close to a high symmetry plane (R point in the cubic phase). 

\begin{figure*}[ht]
\includegraphics[width=1.0\columnwidth]{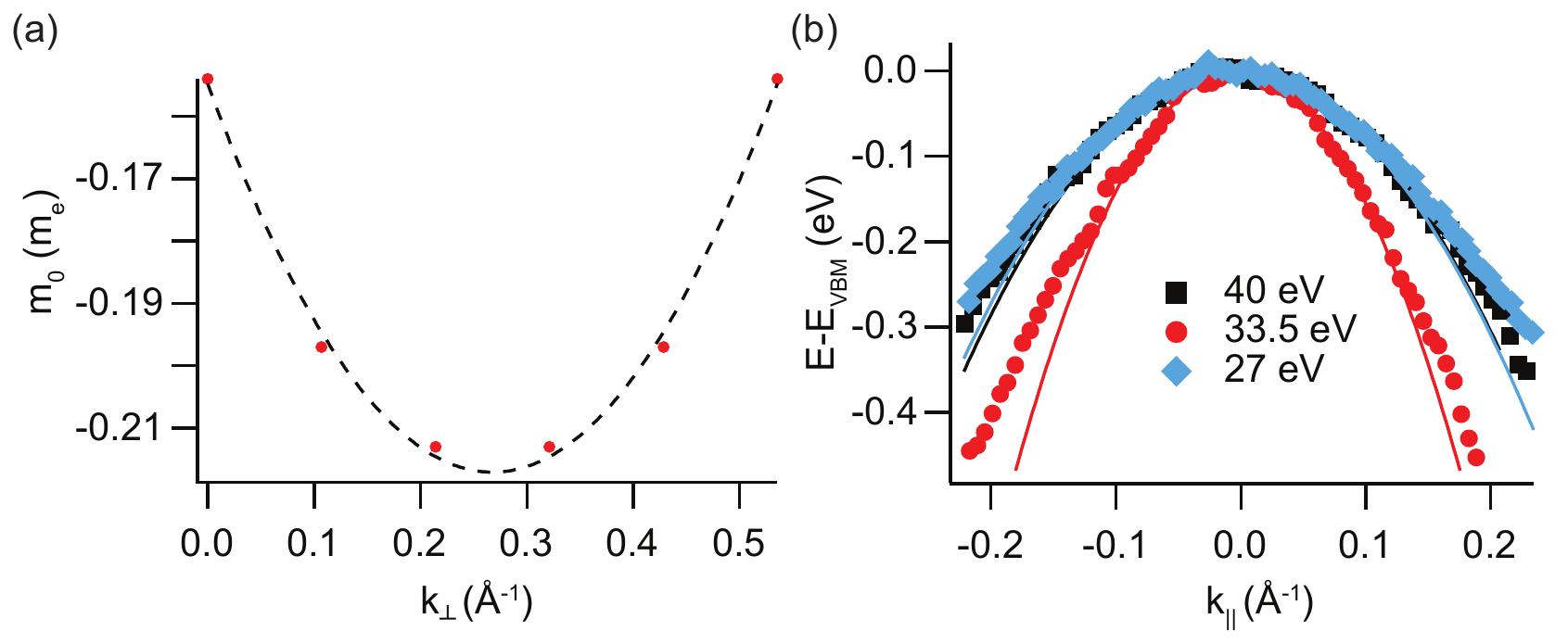}
\caption{\label{fig:kperpmass} (a) Theoretical effective mass extracted from theory (PBE) as a function of k$_{\perp}$ at the $\overline{M}$ point. (b) Experimental band dispersion for three photon energies, parabolic band are fitted to the data, following the method illustrated in the main text. }  
\end{figure*}

The theoretical k$_{\perp}$ evolution of the band dispersion at the $\overline{M}$ point is plotted in Fig. \ref{fig:kperpmass} (a) and predicts a minimum hole effective mass $\approx 0.15 m_{e}$ at $k_{\perp}=0.54$ \AA$^{-1}$. The experimental dispersion is plotted for three photon energies (27, 33.5 and 40 eV) in Fig. \ref{fig:kperpmass} (b): the lighter mass observed at 33.5 eV well agrees with the free electron final state results. 

\subsection{X-ray diffraction characterization}
\begin{figure*}[ht]
\includegraphics[width=0.75\columnwidth]{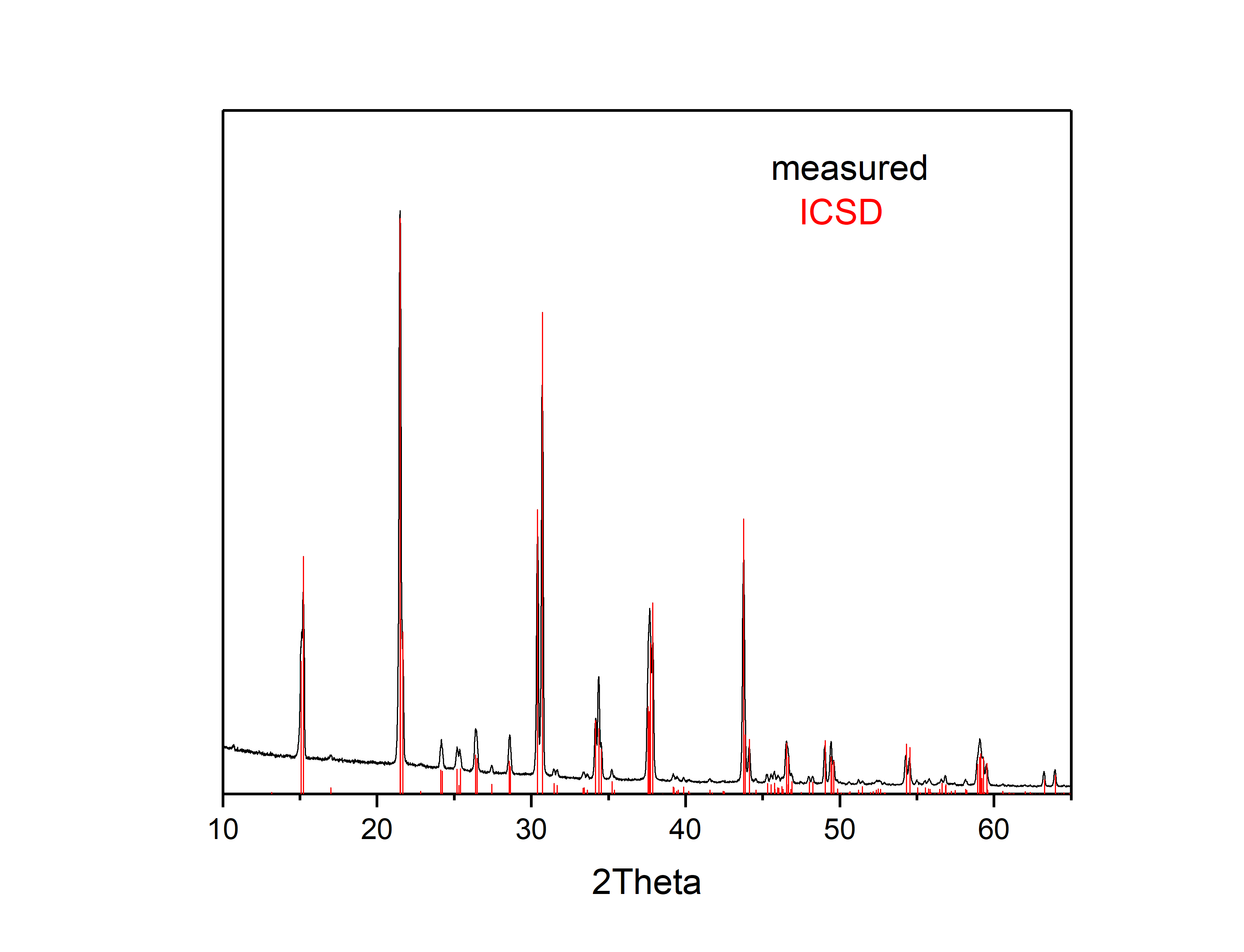}
\caption{\label{fig:XRD} In black: measured powder X-ray diffraction pattern of the samples at room temperature; in red: diffraction pattern from the ICSD database.}  
\end{figure*}
Fig. \ref{fig:XRD} shows the powder X-ray diffraction pattern measured at room temperature. It yields the following crystal cell parameters: a=8.257 \AA, b=8.2160 \AA, c=11.716 \AA, which confirms the orthorhombic symmetry of the crystal, in agreement with previously reported data by Stoupmpos et al. \cite{stoumpos_crystal_2013_SI}.

\section{Theoretical methods}
\subsection{Electronic Structure calculation}
All density functional theory (DFT) calculations were carried out using the 
\textsc{Quantum ESPRESSO} distribution~\cite{giannozzi_quantum_2009_SI, giannozzi_advanced_2017_SI}.
We performed DFT calculation for both the cubic and the orthorhombic phase
(corresponding to the room-temperature stable phase) and set the lattice parameters equal
to the experimental ones~\cite{stoumpos_crystal_2013_SI}.
The electron-ion interactions were modeled using Optimized Norm-Conserving 
Vanderbilt (ONCV) pseudopotentials~\cite{hamann_optimized_2013_SI} as developed by 
Schlipf and Gygi~\cite{schlipf_optimization_2015_SI}. The electronic-structure calculations 
were performed at the generalized Kohn-Sham level using the hybrid functional scheme 
proposed by Heyd, Scuseria and Ernzerhof~\cite{heyd_hybrid_2003_SI,heyd_erratum_2006_SI}
(HSE) for the exchange and correlation energy functional, and including the spin-orbit 
coupling.
An energy cut-off of 80 Ry was used for the plane-wave expansion of the wave-functions
(320 Ry for the charge density) and the Brillouin zone (BZ) was sampled with a 
uniform $\Gamma$-centered mesh of $6\times6\times4$ points ( $6\times 6\times 6$ for the cubic phase).
A reduced (density-) cutoff of 90 Ry and a grid of $3 \times3 \times 2$ points  was
used for the evaluation of the non-local component of the exchange energy and 
potential (the full $6\times 6\times 6$ grid was used for the cubic structure).  

To further check the reliability of the HSE functional, the quasi-particle band structure for the cubic phase 
was also evaluated within the G$_0$W$_0$ approximation using the PBE ground state density and wave-functions as
starting point and including spin-orbit coupling as implemented in the Yambo code~\cite{marini_yambo_2009_SI, sangalli_many-body_2019_SI}. 
Pseudopotentials including all semi-core electrons were used in this case. The parameters used for
the calculation are: 80 Ry plane wave cut-off for the PBE ground state calculation, 15 Ry plane
wave cut-off for the polarizability calculation, 500 bands, 1 Ry  plasmon-pole energy and a
$6\times 6\times 6$ $\Gamma$-centered grid for the BZ integration.

Maximally localized Wannier functions~\cite{marzari_maximally_2012_SI} were computed with the Wannier90 code~\cite{mostofi_wannier90_2008_SI,mostofi_updated_2014_SI} and used to interpolate the HSE (and the G$_0$W$_0$)
band structure on an arbitrary $\mathbf{k}$-point mesh. Interpolated band structure has been used to evaluate
the effective masses, as described in the next section. 

\begin{figure*}[ht]
\includegraphics[width=0.75\columnwidth]{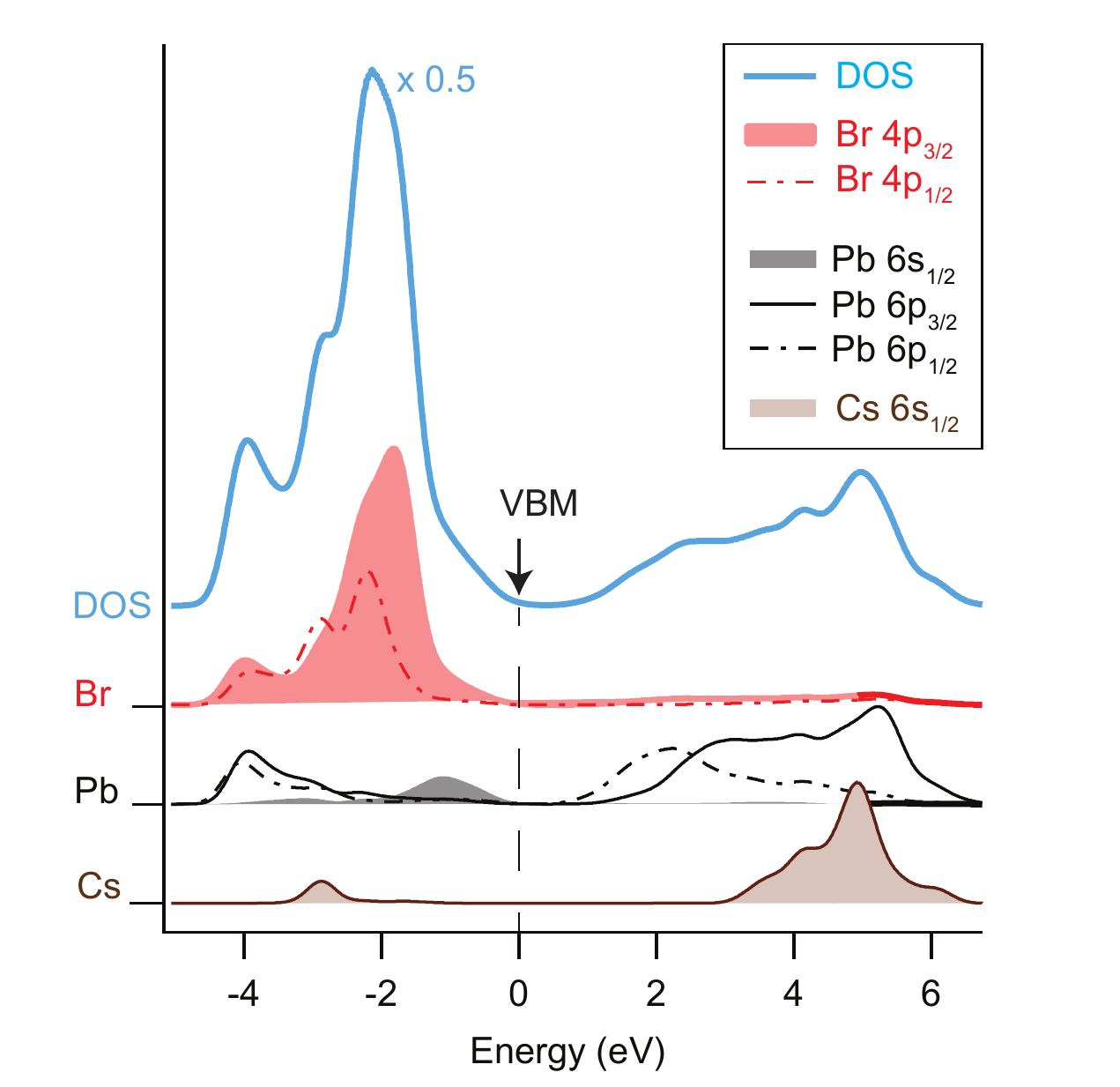}
\caption{\label{fig:PDOS} Total and projected density of states for valence and conduction band. The orbital character is indicated in the legend for the three constituents Br, Pb and Cs. The indicated maximum of the valence band, was in this case set to the onset of the DOS.}  
\end{figure*}

The contributions of electronic states of each individual chemical element of CsPbBr$_3$: Br, Cs, and Pb, to the total density of electronic states (DOS) were calculated and shown in Fig. \ref{fig:PDOS}. 

\subsection{Determination of effective mass from ab-initio DFT bands}
The theoretical effective masses at the top of the valence band (R point for the cubic phase, $\Gamma$ point in the orthorhombic phase) were calculated by evaluating numerically the second derivative of the Wannier-interpolated band structure $\varepsilon (\mathbf{k})$.
\begin{equation}
 [m_0]^{-1}=\frac{1}{\hbar^2}\frac{d^2 \varepsilon(\mathbf{k})}{dk^2}
\end{equation}
For the cubic phase a small step $\Delta k$ has been taken along the [110] direction ($\overline{\Gamma M}$ direction). For the orthorhombic phase we calculated the effective masses along the three crystallographic direction [100], [010], [001]. The $\Delta k$ is reduced until convergence in the second derivative is achieved (typically for $\Delta k \sim 
0.01$ \AA$^{-1}$, see Fig.~\ref{fig:mass_conv}). The converged results, reported in Tab.~\ref{tab:mass_theo} show that 
an improved description of the electronic correlation, i.e. going from PBE to HSE to G$_0$W$_0$ band structure, leads
to an increase of the effective masses. Moreover for the cubic phase we notice that the HSE and G$_0$W$_0$ effective masses 
are quite similar and differ by  $\sim 8$\%, indicating that the HSE functional gives a reasonable description of the
band structure close to the VBM.

\begin{figure}
    \centering
    \includegraphics[width=0.95\columnwidth]{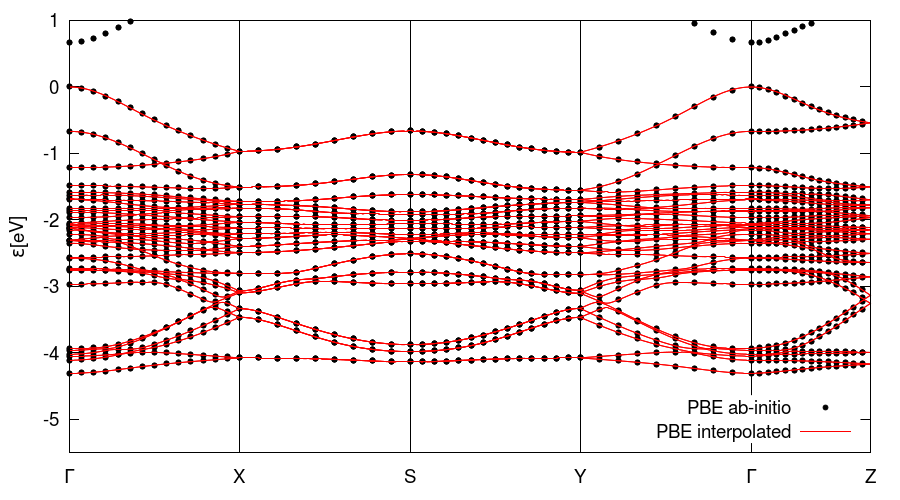}
    \caption{Comparison between the ab-initio and Wannier-interpolated valence band structure for the orthorhombic phase at the PBE level.}
    \label{fig:abinitio_vs_interp}
\end{figure}

\begin{figure}
    \centering
    \includegraphics[width=0.95\columnwidth]{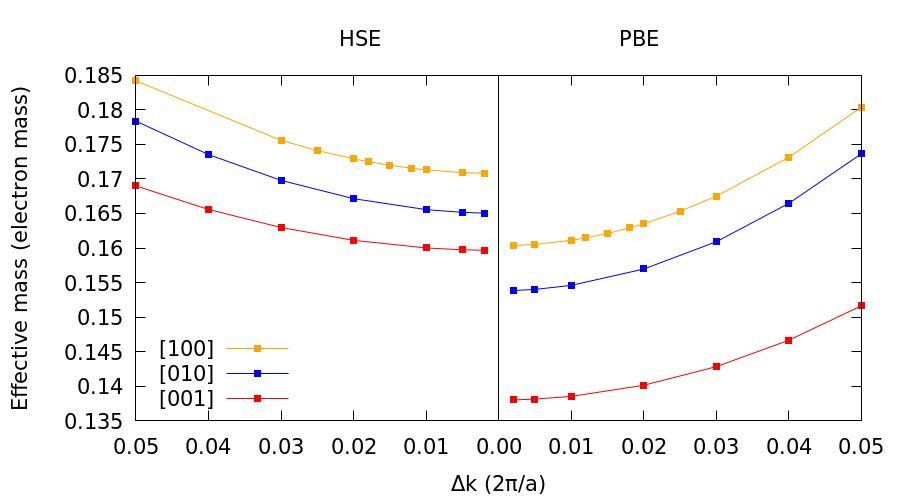}
    \caption{Convergence of the PBE (right panel) and HSE (left panel) effective masses for the orthorhomibc phase with respect to the $\mathbf{k}$-point sampling around the $\Gamma$ point. The distance in the reciprocal space is expressed in unit of $2\pi/a$ with $a$ the 
    lattice parameter along the direction considered.}
    \label{fig:mass_conv}
\end{figure}

Despite the good quality of the Wannier interpolation (see a comparison between the fully ab-initio PBE band structure and the interpolated one in Fig.~\ref{fig:abinitio_vs_interp}), we point out that a small error in the absolute value of the effective masses evaluated from the interpolated bands might still be present. For the PBE functional a direct evaluation of the eigenvalues at any $\mathbf{k}$ point is also possible (this is not the case for the HSE functional). 
A comparison between the PBE effective masses obtained without the interpolation and after the interpolation
is reported in Tab.~\ref{tab:mass_theo}, and reveals a small overestimation of the effective masses ($\sim7$\%). Overall, we are confident that our estimation of the effective masses from the interpolated band structure is correct within 0.01 m$_e$. 

{}

{}

\begin{table}[h]
    \centering
    \begin{tabular}{ |c | c | c | c | }
    \hline 
            &    PBE     &  HSE    &  G$_0$W$_0$ \\
    \hline
    Cubic [110] &    0.123    &  0.148  &   0.139     \\
    \hline
    Ortho [100] & 0.161 (0.149)  &  0.171 &   -- \\
    Ortho [010] & 0.155 (0.142)  &  0.165 &   -- \\
    Ortho [001] & 0.138 (0.128)  &  0.160 &   --  \\
    \hline
    \end{tabular}
    \caption{Hole effective masses at the top of the valence band. The number in parenthesis are the values obtained without the interpolation (only possible for the PBE functional). }
    \label{tab:mass_theo}
\end{table}

\subsection{Electron-phonon interaction}

The electron-LO phonon (longitudinal optical phonons) interaction was accounted for within a multi-phonon Fr\"ohlich
model~\cite{frohlich_electrons_1954_SI, schlipf_carrier_2018_SI}, i.e. assuming parabolic
electronic bands and dispersionless LO phonons, and neglecting acoustic and TO phonons.
The scattering by LO phonons is believed to be the most relevant process for this class of
materials.~\cite{wright_electronphonon_2016_SI, schlipf_carrier_2018_SI}
We obtained the parameters of the model averaging the ab-initio Fr\"ohlich vertex
over $N_{\mathbf{q}}$ = 1000 $\mathbf{q}$ vectors of length 0.001 Bohr$^{-1}$
uniformly distributed around the BZ center. The Fr\"ohlich vertex is 
defined~\cite{vogl_microscopic_1976_SI, verdi_frohlich_2015_SI} as
\begin{equation}
g_{\nu}(\mathbf{q}) = -i\frac{4\pi e^2}{\Omega} \sum_k \sqrt{\frac{\hbar}{2M_k\omega_{\mathbf{q}_{\nu}}}} 
                \frac{ \hat{\mathbf{q}} \cdot Z^*_k \cdot \mathbf{e}_{k_{\nu}}(\mathbf{q}) }{ \hat{\mathbf{q}} \cdot \varepsilon_{\infty} \cdot \hat{\mathbf{q}} }
\end{equation}
where $\Omega$ is the volume of the unit cell, $M_k$ the mass of the atom $k$, $Z^*_k$ the Born effective
charge tensor, $\varepsilon_{\infty}$ the high-frequency dielectric tensor, and $\omega_{\mathbf{q}\nu}$ and 
$\mathbf{e}_{k\nu}(\mathbf{q})$ the eigenvalue and eigenvector associated to the mode $\mathbf{q}{\nu}$. 
All the ingredients above were computed using density functional perturbation theory 
as implemented in the PHONON code of \textsc{Quantum ESPRESSO} and using the 
PBE~\cite{perdew_generalized_1996_SI} functional to account for exchange-correlation effects.
The dynamical matrix has been computed in reciprocal space on a coarse grid of $4\times 4\times 4$ $\mathbf{q}$-point and then interpolated with standard techniques~\cite{baroni_phonons_2001_SI} and with a separate treatment of the long-range dipole-dipole interaction~\cite{gonze_dynamical_1997_SI}. 


In Fig. (4b) of the main text the density of polar coupling~\cite{schlipf_carrier_2018_SI} defined as
\begin{equation}
 \frac{dg^2(\hbar\omega)}{d\omega} = \frac{1}{N_{\mathbf{q}}} \sum_{\mathbf{q}\nu} \delta(\hbar\omega -\hbar\omega_{\mathbf{q}_{\nu}})|g_{\nu}(\mathbf{q})|^2,
\end{equation}
is shown, together with the frequency dependent dielectric function in the infrared region.
The plots highlight that there is one dominant contribution at an average energy of $\hbar\tilde{\omega} = 18.2 $ meV.
The corresponding interaction strength, averaged over the $N_{\mathbf{q}}$ $\mathbf{q}$-points is $|\tilde{g}|^2 =
 3.34\times 10^{-5} $ (eV/\AA)$^2$. 
Following Ref.~\cite{frohlich_electrons_1954_SI}, a dimensionless parameter $\alpha_{\nu}$ can be defined 
for each relevant mode (only one in this case):
\begin{equation}
 \alpha_{\nu} = \frac{\Omega}{4\pi e^2} \frac{ |\tilde{g_{\nu}}|^2 }{(\hbar\tilde{\omega}_{\nu})^2} 
                \left( \frac{2m^*\tilde{\omega}_{\nu}}{\hbar} \right)^{1/2}.
\end{equation}
with $m^*$ the hole effective mass. 
Inserting the effective phonon frequency and interaction strength for the unique relevant LO
phonon found from the analysis above, ad using the HSE effective mass ($m^* =0.171$), we obtain $\alpha=1.82$, which fall into the moderate-coupling regime.

\break
\bibliographystyle{ieeetr}
\bibliography{refs}

\end{document}